# Cybersecurity skills in new graduates: a Philippine perspective

**John Paul P. Miranda[1], Marlon I. Tayag[2], Joel D. Canlas[3]**

[1]College of Computing Studies, Pampanga State University, Mexico, Philippines
[2]School of Computing, Holy Angel University, Angeles City, Philippines
[3]College of Computing Studies, Pampanga State University, Bacolor, Philippines

| Article Info | ABSTRACT |
|---|---|
| *Article history:*<br><br>Received Mar 6, 2025<br>Revised Oct 1, 2025<br>Accepted Nov 4, 2025 | This study investigates the key skills and competencies needed by new cybersecurity graduates in the Philippines for entry-level positions. Using a descriptive cross-sectional research design, it combines analysis of job listings from Philippine online platforms with surveys of students, teachers, and professionals. The aim is to identify required skills and areas needing improvement, highlighting the balance between technical skills and other competencies like ethical conduct, suggesting a shift away from traditional cybersecurity skills towards a more diverse skillset. Furthermore, the results revealed common agreement on the importance of communication, critical thinking, problem-solving, and adaptability skills, albeit with slight variations in their prioritization. It recommends that aspiring cybersecurity professionals develop an inclusive skill set encompassing technical knowledge, soft skills, and personal competencies, with a focus on adaptability, continuous learning, and ethics. Skills such as business acumen are considered less vital for entry-level roles, proposing a preparation strategy that aligns with the changing demands of the cybersecurity industry. |
| *Keywords:*<br><br>Cybersecurity education<br>Employability skills<br>Entry-level qualification<br>Quality education<br>Workforce preparedness | |
| |  |

*Corresponding Author:*

John Paul P. Miranda
College of Computing Studies, Pampanga State University
San Juan, Mexico, Pampanga, Philippines
Email: jppmiranda@pampangastateu.edu.ph

## 1. INTRODUCTION

Cybersecurity is a critical area in today's digital society. However, there remains a global shortage of skilled professionals who can respond to growing security threats. In the Philippines, this shortage is more severe due to limited access to resources, rapid technological adoption, and the increasing number of cyberattacks. Although colleges and universities in the country have introduced programs focused on cybersecurity, there is still a gap between the skills taught in school and the skills required by the industry [1], [2].

Several researchers have already explored ways to reduce the skills gap in cybersecurity [3], focused on measuring the cybersecurity skills of non-IT professionals, while [4] developed models to improve the competencies of students in computing. In [5] examined the changing landscape of cybersecurity education in Europe and highlighted the need to include soft skills. In [6] also looked into how well university curricula in the Philippines match industry expectations. Although these studies provide helpful insights, most of them do not fully examine the Philippine setting or give equal attention to both technical and non-technical skills. Recent global cybersecurity frameworks have also stressed that graduates need a balance of technical skills, risk management, and professional conduct [7], [8] which are still underdeveloped in the Philippines [9].

There are still several unsolved problems. Employers in the Philippines prefer candidates who already have experience, certifications, and strong soft skills. Many new graduates are unable to meet these expectations due to financial limitations, lack of industry exposure, and limited development of non-technical competencies such as communication, adaptability, and ethical judgment [10], [11]. This situation reflects international workforce reports [9], [12]. A security workforce study noted a global shortfall of more than three million professionals [12], while the World Economic Forum's Future of Jobs Report highlighted cybersecurity as an area where both digital and soft skills are increasingly expected [13]. At the regional level, the ASEAN Cybersecurity Cooperation Strategy and ASEAN Digital Masterplan 2025 both call for stronger skills development [14], [15]. Locally, the Department of Information and Communications Technology (DICT) and the Commission on Higher Education (CHED) have introduced policies to strengthen cybersecurity training, although their alignment with international benchmarks is still in progress [2], [16].

This study aims to address these challenges by analyzing the skills and competencies required for entry-level cybersecurity jobs in the Philippines. It combines an analysis of online job listings with a survey of students, teachers, and professionals. Unlike earlier studies, this paper compares the views of multiple stakeholders and identifies the gaps between academic preparation and industry demands. It also highlights skills that may be overlooked in current training programs. The contribution of this study lies in its combined approach: it draws from job market analysis and international frameworks, compares perspectives across groups, and suggests ways to guide local education and policy in line with regional and global trends.

## 2. METHOD

This research employed a cross-sectional, descriptive quantitative approach to examine the varying perceptions of the essential skills required to secure an entry-level position in cybersecurity within the Philippine context. To ascertain the key skills for entry-level cybersecurity roles in the Philippines, this study analyzed popular job listings from prominent Philippine job websites like JobStreet and Indeed Philippines. Focusing on postings from September to November 2023, the research identified frequently mentioned skills, which informed the development of the survey instrument.

The first section of the final questionnaire comprised demographic details, such as age, sex at birth, respondent type, highest education level attained, and specific information relevant to each respondent category (e.g., industry focus for professionals and teachers, school type for students, years of experience for teachers and professionals, and income bracket based on the 2023 Philippine Statistics Authority clustering for professionals and teachers). Among the 176 respondents, 20 were ineligible as they do not possess any knowledge, experience, or background in Cybersecurity. Similar studies have used the same range of respondents to assess both technical and non-technical skills needed in the cybersecurity workforce [17], [18].

As seen in Table 1, out of the 152 respondents, 90 were students, 26 were teachers, and 36 were professionals. The students were enrolled in cybersecurity and related programs at both public and private higher education institutions. Teachers represented colleges and universities offering computing and engineering courses, while professionals were distributed across IT services, software development, government, and non-profit organizations. The age ranges for the respondents were 18 to 24 years for students, 31 to 47 years for teachers, and 27 to 44 years for professionals, with the professional group showing the highest variability. Respondents across all groups were predominantly male. In terms of educational background, teachers reported higher qualifications than professionals, particularly with respect to postgraduate degrees.

Regarding the nature of work, teachers were mainly engaged in education and training, while professionals indicated involvement in diverse sectors, though primarily in the private sector. Teachers were almost evenly split between public and private institutions. Work experience also varied, with teachers reporting between 5 and 21 years of service and professionals between 3 and 20 years. Income levels reflected these differences: professionals were spread across a wider set of income brackets, with most earning between ₱48,328 and ₱84,574, whereas teachers' incomes were more concentrated between ₱24,164 and ₱48,328.

The second section of the survey contained two key questions about the essential skills for securing an entry-level cybersecurity job. A total of 36 vital job-related skills in cybersecurity were pinpointed through this analysis in Table 2 (see Appendix). Respondents rated each skill using a 10-point Likert-type scale, where 1 indicated "very useful," and 10 indicated "least useful." They also provided a second rating to indicate whether the skill required further development. This dual structure allowed the survey to capture both perceived importance and training gaps. The instrument was subjected to validity and reliability testing. For validity, two professors with expertise in cybersecurity education reviewed the instrument to ensure its appropriateness for the study, and it passed the inter-rater reliability test to assess its consistency. The study utilized Google Forms for data collection, and each respondent was queried about their knowledge, experience, or background in cybersecurity in any capacity.

Table 1. Demographic details

| Demographic | Student | Frequency teacher | Professional |
|---|---|---|---|
| Age (mean±SD) | (21.78±3.19) | (39.42±7.59) | (36.14±8.53) |
| Sex at birth | | | |
| − Male | 65 | 15 | 28 |
| − Female | 24 | 11 | 8 |
| − Prefer not to say | 1 | - | - |
| Highest educational | | | |
| − Secondary | 90 | - | - |
| − Bachelor | - | 6 | 23 |
| − Master | - | 13 | 11 |
| − Doctorate | - | 7 | 2 |
| Primary nature of work | | | |
| − Information technology (IT) and software development | - | 2 | 18 |
| − Business process outsourcing (BPO)/call centers | - | - | 1 |
| − Manufacturing and production | - | - | 3 |
| − Engineering | - | - | 1 |
| − Education and training | - | 24 | 9 |
| − Government and non-profit organizations | - | - | 4 |
| Type of work | | | |
| − Public | - | 15 | 13 |
| − Private | - | 11 | 23 |
| No. of years working (mean±SD) | - | (13.62±7.77) | (11.61±8.45) |
| Income cluster | | | |
| − Less than ₱12,082 | - | - | 1 |
| − Between ₱12,082 and ₱24,164 | - | 2 | 5 |
| − Between ₱24,164 and ₱48,328 | - | 18 | 8 |
| − Between ₱48,328 and ₱84,574 | - | 6 | 12 |
| − Between ₱84,574 and ₱144,984 | - | - | 6 |
| − Between ₱144,984 and ₱241,640 | - | - | 4 |

## 3. RESULTS AND DISCUSSION

Table 3 provides comparative results of the analysis of the perceived importance of most needed skills and competencies for securing an entry-level job in cybersecurity, as assessed by students, teachers, and professionals. The table reveals that skills such as communication, critical thinking, problem-solving, teamwork, personal values, ethical mindset, professionalism, continuous learning, risk assessment, adaptability, self-motivation, resilience and stress management, personal accountability, and self-confidence are deemed highly necessary. These skills were primarily rated with a median of 1 or 1.5, indicating a strong consensus between teachers and professionals regarding their importance. While a broad agreement exists across all participant groups on the importance of a diverse set of skills and competencies, professionals emphasized the critical nature of 22 out of 36 skills and competencies, marking them with the highest importance. This highlights the view of the professionals that these skills are crucial for success in entry-level roles [19].

The assessments of teachers varied slightly, with certain skills such as critical thinking, problem-solving, teamwork, personal values, ethical mindset, professionalism, continuous learning, personal accountability, and conflict resolution being highlighted as particularly important [19]. In contrast, students generally assigned a rating of 2 to the majority of skills, recognizing their importance but with a marginally lower sense of urgency compared to teachers and professionals. This difference in perception may stem from the students' relative inexperience and impending entry into the job market [20]. Interestingly, students placed a high value on cybersecurity certifications across all skills and competencies.

At the same time, the results also emphasized the critical role of adaptability, continuous learning, resilience and stress management, and personal accountability, highlighting these as universally important across all groups. This reflects the dynamic nature of cybersecurity work, emphasizing not just technical skills but also the necessity for ongoing learning, stress management, and adaptability in a rapidly evolving field [11]. It is interesting to note that skills such as cybersecurity certifications, incident handling and forensic skills, and financial literacy were ranked lower by professionals [21], which suggests these areas might be more pertinent at later career stages rather than at the entry level. This insight points to a nuanced understanding of skill importance, distinguishing between foundational competencies and those that become more relevant as one's career progresses.

Table 4 provides a ranked assessment of skills and competencies, assessing their perceived urgency for improvement across different experience levels within the same groups. The majority of the skills are uniformly ranked and have a median of 2 across all three groups, which indicates a widespread agreement on which areas demand the most attention. However, certain skills exhibit minor variances in perception among the groups. For instance, critical thinking, information technology/cybersecurity skills, planning, and problem-solving are deemed more critical by teachers [22], who assign them a median rank of 1, compared to students and professionals, who generally assign a median rank of 2. This difference likely mirrors the belief of teachers in the essential role these skills play in fostering effective learning and their application within professional settings. Conversely, skills such as cross-disciplinary knowledge and assertiveness received a slightly lower urgency rating of 2.5 from students, suggesting they view these areas as less immediate in their need for attention compared to others. Similarly, cybersecurity certifications, entrepreneurial skills, and financial literacy are ranked at 3 by professionals, indicating that, from their perspective, while still important, these areas do not require as immediate attention as others.

Table 3. Median rank of the most needed skills to land an entry-level job

| Skill/competencies | Students | Teachers | Professionals |
|---|---|---|---|
| Communication | 2 | 1.5 | 1 |
| Critical thinking | 2 | 1 | 1 |
| Entrepreneurial | 2 | 2.5 | 1 |
| Human relations | 2 | 2 | 1 |
| Information technology/cybersecurity skills | 2 | 1.5 | 1 |
| Planning | 2 | 2 | 1 |
| Problem-solving | 2 | 1 | 1 |
| Research | 2 | 2 | 1 |
| Teamwork | 2 | 1 | 1 |
| Personal values | 2 | 1 | 1 |
| Ethical mindset | 2 | 1 | 1 |
| Flexibility | 2 | 2 | 1 |
| Professionalism | 2 | 1 | 1 |
| Attention to detail | 2 | 2 | 1 |
| Risk assessment | 2 | 1.5 | 1 |
| Adaptability | 2 | 1.5 | 1 |
| Continuous learning | 2 | 1 | 1 |
| Self-motivated | 2 | 1.5 | 1 |
| Resilience and stress management | 2 | 1.5 | 1 |
| Personal accountability | 2 | 1 | 1 |
| Self-confidence | 2 | 1.5 | 1 |
| Strategic thinking | 2 | 2 | 1 |
| Leadership | 2 | 2 | 2 |
| Legal and compliance knowledge | 2 | 2 | 2 |
| Training/coaching | 2 | 1.5 | 2 |
| Time management | 2 | 2 | 2 |
| Cross-disciplinary knowledge | 2 | 2 | 2 |
| Vendor and tool familiarity | 2 | 2 | 2 |
| Business acumen | 2 | 2 | 2 |
| Incident handling and forensic skills | 2 | 2 | 2 |
| Conflict resolution | 2 | 1 | 2 |
| Assertiveness | 2 | 2 | 2 |
| Goal setting | 2 | 2 | 2 |
| Financial literacy | 3 | 2 | 3 |
| Negotiation | 2 | 2 | 3 |
| Cybersecurity certifications | 1.5 | 2 | 3 |

The Kruskal-Wallis H test was utilized to evaluate if the importance rankings assigned to various skills by respondents significantly varied for obtaining an entry-level cybersecurity position (see Table 5). The results indicated that certain skills specifically, critical thinking (H=7.654; p=0.02), problem-solving (H=7.712; p=0.02), attention to detail (H=7.049; p=0.03), and personal accountability (H=7.295; p=0.03) were significantly ranked differently, highlighting their perceived importance for entry-level roles in cybersecurity. Furthermore, a group of skills approached the significance threshold (p-values near 0.05), suggesting they could be considered nearly significant. These include teamwork (H=5.811; p=0.06), flexibility (H=5.685; p=0.06), professionalism (H=5.511; p=0.06), risk assessment (H=4.888; p=0.09), personal values (H=4.784; p=0.09), ethical mindset (H=4.538; p=0.10), and communication (H=4.595; p=0.10). While these skills did not meet the conventional significance threshold, they show a trend that may reach statistical significance with an increased sample size or under different circumstances, suggesting variability in their perceived importance for entry-level cybersecurity positions.

Conversely, the study also found a general agreement on the importance of most skills, indicating consistent perceptions among respondents about their necessity for securing an entry-level job in cybersecurity. Skills with the least significance, such as business acumen (H=0.37; p=0.83) and financial literacy (H=0.18; p=0.91), were seen as not differing significantly among respondents. The differing perspectives on the importance of specific skills among students, teachers, and professionals, especially those with significant scores which imply several critical aspects for cybersecurity education and training [23]. This variance necessitates adjustments in curriculum development to better align educational programs with industry needs, highlighting a gap in career preparedness that educational institutions need to address. The results suggest that curricula may place greater emphasis on soft skills, case-based problem-solving, and those related to ethical judgment. Internship opportunities and accessible certification pathways can also help students connect academic preparation with workplace expectations. At the policy level, agencies such as DICT and CHED may use these insights to refine training standards and curricula. Closer collaboration between universities and industry can also support a stronger talent pipeline. This also points to the studies where a need for ongoing professional development and training tailored to the most valued skills, as well as enhanced mentorship for students and entry-level professionals [24], [25].

Table 4. Median rank of skills needing the most attention based on experience

| Skill/competencies | Students | Teachers | Professionals |
|---|---|---|---|
| Communication | 2 | 2 | 2 |
| Critical thinking | 2 | 1 | 2 |
| Human relations | 2 | 2 | 2 |
| Information technology/cybersecurity skills | 2 | 1 | 2 |
| Leadership | 2 | 2 | 2 |
| Planning | 2 | 1 | 2 |
| Problem-solving | 2 | 1 | 2 |
| Research | 2 | 2 | 2 |
| Training/coaching | 2 | 2 | 2 |
| Teamwork | 2 | 2 | 2 |
| Personal values | 2 | 2 | 2 |
| Ethical mindset | 2 | 2 | 2 |
| Flexibility | 2 | 2 | 2 |
| Professionalism | 2 | 2 | 2 |
| Legal and compliance knowledge | 2 | 2 | 2 |
| Time management | 2 | 2 | 2 |
| Attention to detail | 2 | 2 | 2 |
| Risk assessment | 2 | 2 | 2 |
| Adaptability | 2 | 2 | 2 |
| Cross-disciplinary knowledge | 2.5 | 2 | 2 |
| Vendor and tool familiarity | 2 | 2 | 2 |
| Continuous learning | 2 | 2 | 2 |
| Business acumen | 2 | 2 | 2 |
| Incident handling and forensic skills | 2 | 2 | 2 |
| Self-motivated | 2 | 2 | 2 |
| Resilience and stress management | 2 | 2 | 2 |
| Personal accountability | 2 | 2 | 2 |
| Conflict resolution | 2 | 2 | 2 |
| Assertiveness | 2.5 | 2 | 2 |
| Self-confidence | 2 | 2 | 2 |
| Goal setting | 2 | 2 | 2 |
| Strategic thinking | 2 | 2 | 2 |
| Negotiation | 2 | 2 | 2 |
| Cybersecurity certifications | 2 | 2 | 3 |
| Entrepreneurial | 2 | 2 | 3 |
| Financial literacy | 2 | 2 | 3 |

Furthermore, this situation underlines the importance of collaboration among educational institutions, industry professionals, and policymakers to ensure that the training and education provided are relevant and effective in preparing a skilled workforce ready to tackle the cybersecurity challenges of tomorrow [26], [27]. Addressing these discrepancies is vital for developing a comprehensive strategy that bridges the gap between educational outcomes and industry expectations. These may enhance the overall efficacy of cybersecurity practices and workforce readiness.

Table 6 presents the results of the Kruskal-Wallis H test, which was also conducted to determine whether the importance rankings of skills requiring the most attention, based on experience, varied among respondents. The analysis revealed that problem-solving (H=6.899; p=0.03) and incident handling and forensic skills (H=8.441; p=0.02) exhibited statistically significant differences in their rankings. This variation suggested that the importance attached to these skills differed particularly across experience levels, potentially indicating that more professionals might have prioritized these areas more than their less experienced counterparts, or vice versa. Such differences highlighted the necessity for education and training programs in cybersecurity to be adaptable, catering to the diverse perceptions of skill importance identified among different groups.

Table 5. Test of difference in ranking the most needed skills to land an entry-level job

| Skill/competencies | Kruskall-Wallis $H$ test | Significance |
|---|---|---|
| Communication | 4.595 | 0.10 |
| Critical thinking | 7.654 | 0.02 |
| Entrepreneurial | 2.194 | 0.33 |
| Human relations | 2.929 | 0.23 |
| Information technology/cybersecurity skills | 4.034 | 0.13 |
| Leadership | 1.044 | 0.59 |
| Planning | 1.115 | 0.57 |
| Problem-solving | 7.712 | 0.02 |
| Research | 1.213 | 0.55 |
| Training/coaching | 1.788 | 0.41 |
| Teamwork | 5.811 | 0.06 |
| Personal values | 4.784 | 0.09 |
| Ethical mindset | 4.538 | 0.10 |
| Flexibility | 5.685 | 0.06 |
| Professionalism | 5.511 | 0.06 |
| Legal and compliance knowledge | 0.963 | 0.62 |
| Time management | 0.782 | 0.68 |
| Attention to detail | 7.049 | 0.03 |
| Risk assessment | 4.888 | 0.09 |
| Adaptability | 2.152 | 0.34 |
| Cybersecurity certifications | 3.502 | 0.17 |
| Cross-disciplinary knowledge | 3.268 | 0.20 |
| Vendor and tool familiarity | 2.15 | 0.34 |
| Continuous learning | 2.725 | 0.26 |
| Business acumen | 0.37 | 0.83 |
| Incident handling and forensic skills | 2.443 | 0.30 |
| Self-motivated | 5.356 | 0.07 |
| Resilience and stress management | 3.698 | 0.16 |
| Personal accountability | 7.295 | 0.03 |
| Conflict resolution | 3.88 | 0.14 |
| Assertiveness | 0.868 | 0.65 |
| Self-confidence | 4.091 | 0.13 |
| Goal setting | 0.939 | 0.63 |
| Financial literacy | 0.18 | 0.91 |
| Strategic thinking | 1.669 | 0.43 |
| Negotiation | 0.77 | 0.68 |

The divergence in how problem-solving and incident handling, and forensic skills were prioritized across groups hinted at underlying differences in practical exposure, immediate relevance, and perceived future importance. Professionals, often on the frontline of cybersecurity defense, likely recognized the critical nature of these skills in addressing real-world challenges, while students may not have fully grasped their complexity or immediate applicability without practical experience. Additionally, skills such as information technology/cybersecurity skills (H=5.741; p=0.06), planning (H=4.521; p=0.10), risk assessment (H=4.926; p=0.09), cross-disciplinary knowledge (H=4.513; p=0.11), and continuous learning (H=4.498; p=0.11) showed p-values nearing the threshold of significance. Although these did not reach statistical significance, the trend suggested a perspective on the importance of these skills, influenced by the level of experience of the respondents. This emerging disagreement warranted further exploration, reflecting the dynamic nature of the cybersecurity field and the imperative for ongoing adaptation and learning across all experience levels.

The evidence of distinct perspectives on critical skills among the respondents in cybersecurity pointed to the need for dynamic, responsive educational frameworks [28]. These frameworks should not only bridge the gap between theory and practice but also cater to the evolving demands of the cybersecurity landscape. Tailoring education and training programs to accommodate these varied perceptions could help ensure that all entering the field are well-equipped to tackle its challenges. Moreover, fostering a dialogue between academia and industry could enhance the alignment of educational outcomes with real-world needs that will help ensure that graduates will be ready to contribute from the outset of their careers.

Table 6. Test of difference in ranking the skills needing the most attention based on experience

| Skill/competencies | Kruskall-Wallis H test | Significance |
|---|---|---|
| Communication | 0.931 | 0.63 |
| Critical thinking | 3.836 | 0.15 |
| Entrepreneurial | 3.882 | 0.14 |
| Human relations | 1.89 | 0.39 |
| Information technology/cybersecurity skills | 5.741 | 0.06 |
| Leadership | 3.609 | 0.17 |
| Planning | 4.521 | 0.10 |
| Problem-solving | 6.899 | 0.03 |
| Research | 2.601 | 0.27 |
| Training/coaching | 3.03 | 0.22 |
| Teamwork | 1.55 | 0.46 |
| Personal values | 2.85 | 0.24 |
| Ethical mindset | 3.481 | 0.18 |
| Flexibility | 0.872 | 0.65 |
| Professionalism | 2.225 | 0.33 |
| Legal and compliance knowledge | 1.089 | 0.58 |
| Time management | 0.224 | 0.89 |
| Attention to detail | 3.468 | 0.18 |
| Risk assessment | 4.926 | 0.09 |
| Adaptability | 1.845 | 0.40 |
| Cybersecurity certifications | 1.522 | 0.47 |
| Cross-disciplinary knowledge | 4.513 | 0.11 |
| Vendor and tool familiarity | 0.955 | 0.62 |
| Continuous learning | 4.498 | 0.11 |
| Business acumen | 0.871 | 0.65 |
| Incident handling and forensic skills | 8.441 | 0.02 |
| Self-motivated | 3.218 | 0.20 |
| Resilience and stress management | 2.251 | 0.32 |
| Personal accountability | 4.258 | 0.12 |
| Conflict resolution | 2.854 | 0.24 |
| Assertiveness | 4.286 | 0.12 |
| Self-confidence | 1.195 | 0.55 |
| Goal setting | 2.489 | 0.29 |
| Financial literacy | 1.991 | 0.37 |
| Strategic thinking | 3.444 | 0.18 |
| Negotiation | 2.464 | 0.29 |

These results provide important insights into the current state of cybersecurity education in the Philippines and how it compares with global trends. While all groups in the study acknowledged the value of key skills, students rated many of them with slightly lower urgency. This difference may reflect limited exposure to actual workplace conditions. In the Philippines, where many institutions are still developing their cybersecurity programs, this result shows the need to strengthen the connection between what is taught in schools and what is expected in professional environments. Similar patterns have been reported in other countries, where employers are now placing greater importance on soft skills such as critical thinking, adaptability, and ethical judgment, along with technical competence.

The results of this study are useful for educators, curriculum developers, and government agencies that aim to improve cybersecurity training. Institutions in the Philippines can use this evidence to enhance programs by including more practical experiences, improving soft skill instruction, and encouraging collaboration with industry. These actions can help students become more prepared for entry-level jobs. At the national level, the findings can support policies that focus on developing a stronger and more adaptable cybersecurity workforce. Globally, this study contributes to the broader understanding of how to align cybersecurity education with the needs of the digital economy. Future research can build on these results by exploring how graduates apply these skills in real work settings and how educational strategies can respond to changes in the global cybersecurity landscape.

## 4. CONCLUSION

This study identifies the wide range of skills needed to succeed in cybersecurity and shows that professionals, teachers, and students agree on the value of communication, critical thinking, problem-solving, and ethical awareness, with professionals emphasizing adaptability, continuous learning, and resilience as core requirements. Teachers rated critical thinking, problem-solving, and professionalism as especially important, while students recognized these skills but with less urgency, and professionals viewed traditional cybersecurity skills like certifications and forensic techniques as less essential for entry-level positions. The results point to a changing landscape where technical ability must align with soft skills and personal competencies, offering aspiring practitioners a clearer direction for developing a balanced skill set. The study remains limited by its modest and region-specific sample, reliance on self-reports, and a short three-month window for job listing analysis, which may narrow the scope of interpretation, yet these constraints open space for future research that may refine skill profiles, align findings with labor market data, explore demographic differences, examine specific soft and technical skill gaps, and consider longitudinal and AI-supported approaches to strengthen evidence for educational and policy development.

## ACKNOWLEDGMENTS
This study was supported by the individual affiliations of the authors.

## FUNDING INFORMATION
The authors declare no funding involved.

## AUTHOR CONTRIBUTIONS STATEMENT

This journal uses the Contributor Roles Taxonomy (CRediT) to recognize individual author contributions, reduce authorship disputes, and facilitate collaboration.

| Name of Author | C | M | So | Va | Fo | I | R | D | O | E | Vi | Su | P | Fu |
|---|---|---|---|---|---|---|---|---|---|---|---|---|---|---|
| John Paul P. Miranda | ✓ | ✓ | | ✓ | ✓ | ✓ | ✓ | ✓ | ✓ | ✓ | | ✓ | ✓ | ✓ |
| Marlon I. Tayag | ✓ | ✓ | | ✓ | ✓ | ✓ | ✓ | ✓ | ✓ | ✓ | | ✓ | ✓ | ✓ |
| Joel D. Canlas | ✓ | ✓ | ✓ | | ✓ | ✓ | ✓ | | ✓ | ✓ | ✓ | | | ✓ |

| | | | | | | |
|---|---|---|---|---|---|---|
| C : **C**onceptualization | | I : **I**nvestigation | | Vi : **Vi**sualization | | |
| M : **M**ethodology | | R : **R**esources | | Su : **Su**pervision | | |
| So : **So**ftware | | D : **D**ata Curation | | P : **P**roject administration | | |
| Va : **Va**lidation | | O : Writing - **O**riginal Draft | | Fu : **Fu**nding acquisition | | |
| Fo : **Fo**rmal analysis | | E : Writing - Review & **E**diting | | | | |

## CONFLICT OF INTEREST STATEMENT
Authors state no conflict of interest.

## INFORMED CONSENT
We have obtained informed consent from all individuals included in this study.

## ETHICAL APPROVAL
The research complied with the tenets of the Belmont Report, Helsinki Declaration, and the Philippine Data Privacy Act of 2012. This research has been approved by the authors' institutional review board.

## DATA AVAILABILITY
The data that support the findings of this study are available from the corresponding author, [JPM], upon reasonable request.

## APPENDIX

Table 2. Key skills and competencies

| Skills/competencies | Definition | Basis |
| --- | --- | --- |
| Communication | The ability to convey information effectively and clearly to team members, clients, and stakeholders, both verbally and in writing, in a cybersecurity context. | [29], [30] |
| Critical thinking | The ability to analyze and evaluate information objectively, identify vulnerabilities, and devise effective solutions in a cybersecurity context. | [31], [32] |
| Entrepreneurial | The ability to think creatively, take initiative, and adapt quickly to emerging challenges and opportunities within the cybersecurity field. | [33], [34] |

Table 2. Key skills and competencies *(continued)*

| Skills/competencies | Definition | Basis |
|---|---|---|
| Human relations | The ability to work effectively with others, understand their perspectives, and collaborate in a team-oriented environment in the context of cybersecurity. | [29], [35] |
| information technology/ cybersecurity skills | Proficiency in utilizing various hardware, software, and networking tools relevant to the field of cybersecurity. | [31], [32], [36] |
| leadership | The ability to guide and inspire a team, make informed decisions, and take charge in managing cybersecurity projects or initiatives | [29], [30], [37] |
| Planning | The ability to organize tasks, set priorities, and create strategic roadmaps for achieving cybersecurity objectives. | [38], [39] |
| Problem-solving | The ability to identify, analyze, and resolve complex issues or security breaches in a cybersecurity context. | [40], [41] |
| Research | The ability to gather and analyze information from various sources to stay updated on the latest cyber threats, vulnerabilities, and mitigation strategies. | [42] |
| Training/coaching | The ability to educate and mentor team members or end-users on cybersecurity best practices and protocols. | [43], [44] |
| Teamwork | The capacity to collaborate effectively with colleagues, share knowledge, and contribute to group efforts in achieving cybersecurity goals. | [29], [32] |
| Personal values | The set of principles and beliefs that guide one's behavior and decision-making, particularly in the context of ethical considerations within cybersecurity. | [35], [45] |
| Ethical mindset | Adherence to a code of conduct that upholds moral and professional standards in the practice of cybersecurity. | [29], [46] |
| Flexibility | The ability to adapt to evolving technologies, methodologies, and threats within the dynamic field of cybersecurity. | [29], [46] |
| Professionalism | Demonstrating a high level of integrity, accountability, and respect for confidentiality while working in a cybersecurity role. | [35], [45] |
| Legal and compliance knowledge | Understanding and adhering to the laws, regulations, and industry standards relevant to cybersecurity practices. | [31], [42] |
| Time management | The skill of effectively allocating and utilizing time to prioritize tasks and meet deadlines in a cybersecurity context. | [47], [48] |
| Attention to detail | The ability to thoroughly review and assess information, ensuring accuracy and identifying potential security risks or vulnerabilities. | [35], [49] |
| Risk assessment | The capability to evaluate and quantify potential cybersecurity threats and vulnerabilities, enabling informed decision-making on risk mitigation. | [30], [31], [42], [50] |
| Adaptability | The capacity to quickly adjust strategies and tactics in response to evolving cybersecurity threats or changes in the environment. | [37], [51], [52] |
| Cybersecurity certifications | Recognized credentials demonstrating proficiency and expertise in specific areas of cybersecurity. | [36], [50] |
| Cross-disciplinary Knowledge | Familiarity with related fields (such as networking, systems administration, or cryptography) that complement and enhance cybersecurity expertise. | [53], [54] |
| Vendor and tool familiarity | Knowledge of specific cybersecurity tools and technologies commonly used in the industry. | [31] |
| Continuous learning | A commitment to staying updated with the latest developments, trends, and best practices in the field of cybersecurity. | [29], [32] |
| Business acumen | Understanding the business context in which cybersecurity operates, including budgeting, resource allocation, and alignment with organizational goals. | [30], [31] |
| Incident handling and forensic skills | Proficiency in responding to security incidents, conducting investigations, and preserving digital evidence for analysis. | [30], [31], [42] |
| Self-motivated | Taking the initiative to improve skills, stay informed about emerging threats, and proactively address potential vulnerabilities. | [32] |
| Resilience and stress management | Building the capacity to handle high-pressure situations, such as responding to security incidents, in a calm and effective manner. | [55], [56] |
| Personal accountability | Taking responsibility for one's actions and decisions, especially in situations where security incidents occur. | [37] |
| Conflict resolution | The ability to manage and resolve conflicts, both internally (within oneself) and externally (with others). | [57] |
| Assertiveness | The ability to express one's needs, opinions, and feelings in a clear, respectful, and direct manner. | [58] |
| Self-confidence | Having faith in one's own abilities, decisions, and judgments. | [29] |
| Goal setting | The skill of setting clear, achievable objectives, and creating a plan to work towards them. | [46] |
| Financial literacy | The ability to manage personal finances, encompassing budgeting, saving, investing, and understanding basic economic concepts, demonstrates an understanding of budgeting and resource allocation, which proves valuable in roles involving the management of cybersecurity projects or initiatives. | [59], [60] |
| Strategic thinking | The ability to assess situations, anticipate future trends, and formulate effective plans or strategies to achieve specific goals and objectives, which in cybersecurity involves devising comprehensive security strategies, anticipating potential threats, and implementing proactive measures to safeguard information systems and networks. | [30], [42] |
| Negotiation | The ability to articulate persuasively, find common ground, and reach mutually satisfactory agreements between parties with differing interests or viewpoints. | [30], [61] |

## BIOGRAPHIES OF AUTHORS

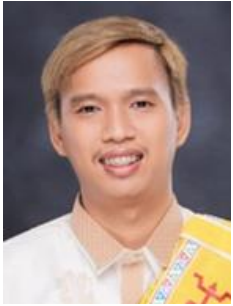 **John Paul P. Miranda** is an Associate Professor and the international linkages and partnerships project head for the Office for International Partnerships and Programs at Pampanga State University. His publications are indexed in Scopus, Web of Science, and IEEE databases. He is a proud member of the National Research Council of the Philippines attached agency to the Department of Science and Technology, which is an advisory body to the Philippine Government on matters of national interest. His area of interest in publications is related to data analytics, educational technology, and application development. He can be contacted at email: jppmiranda@pampangastateu.edu.ph.

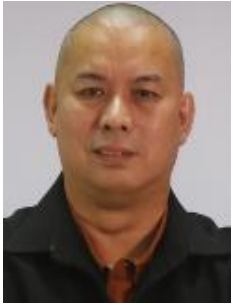 **Marlon I. Tayag** is a full-time Associate Professor III at Holy Angel University and teaches Cyber Security subjects on Ethical Hacking, Digital Forensics, and IoT Security. He earned the degree of Doctor in Information Technology from St. Linus University in 2015 and is currently taking up a Doctor of Philosophy in Computer Science at Technological Institute of the Philippines, Manila. He is a Certified Ethical Hacker, Penetration Tester, Cisco Certified Network Associate, Microsoft Certified Professional, and Microsoft Certified Educator. He can be contacted at email: mtayag@hau.edu.ph.

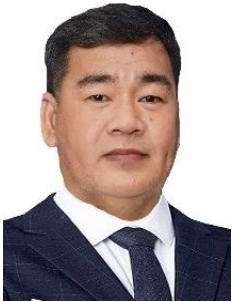 **Joel D. Canlas** is an Associate Professor at Pampanga State University. He is the Dean of the College of Computing Studies at Don Honorio Ventura State University-Main Campus. He can be contacted at email: jdcanlas@pampangastateu.edu.ph.